\documentclass[prb,twocolumn,superscriptaddress,showpacs,floatfix]{revtex4}

\usepackage{epsfig,amsmath,amssymb,latexsym}

\usepackage{subfigure}
\usepackage[normalem]{ulem}
\usepackage{cancel}

\begin{document}

\title{Electronic structure of vacancy-ordered iron-selenide K$_{0.5}$Fe$_{1.75}$Se$_2$} \preprint{1}

\author{Chao Cao}
 \email[E-mail address: ]{ccao@hznu.edu.cn}
 \affiliation{Condensed Matter Group,
  Department of Physics, Hangzhou Normal University, Hangzhou 310036, China}
\author{Fuchun Zhang}
  \email[E-mail address: ]{fuchun@hku.hk}
  \affiliation{Department of Physics,
  Zhejiang University, Hangzhou 310027, China}

\date{Jan. 27, 2013}

\begin{abstract}
The electronic structure of the vacancy-ordered K$_{0.5}$Fe$_{1.75}$Se$_2$ iron-selenide compound (278 phase) is studied using the first-principles density functional method. The ground state of the 278 phase is stripe-like antiferromagnetic, and its bare electron susceptibility shows a large peak around $(\pi, \pi)$ in the folded Brillouin zone. Near Fermi level, the density of states are dominated by the Fe-3d orbitals, and both electron-like and hole-like Fermi surfaces appear in the Brillouin zone. Unfolded band structure shows limited similarities to a hole doped 122 phase. With 0.1e electron doping, the susceptibility peak is quickly suppressed and broadened; while the two-dimensionality of the electron-like Fermi surfaces are greatly enhanced, resulting in a better nesting behavior. Our study should be relevant to the recently reported superconducting phase K$_{0.5+x}$Fe$_{1.75+y}$Se$_2$ with both $x$ and $y$ very tiny.
\end{abstract}

\maketitle

{\it Introduction-}
  The discovery of vacancy-ordered iron-selenides\cite{PhysRevB.82.180520,famous_epl,245_dft,234_dft} has greatly renowned  the study of iron-based superconductors\cite{LOFA_JACS}. These compounds share a general chemical formula of $A_y$Fe$_{2-x}$Se$_2$, and their crystal structures are close to BaFe$_2$As$_2$. However, they differ from the latter because the iron sites are not fully occupied in the selenides, and the iron-vacancies may form ordered structure at certain $x$. For example, in the vacancy-ordered K$_{0.8}$Fe$_{1.6}$Se$_2$ (or K$_2$Fe$_4$Se$_5$, 245 phase), the iron-vacancies form a superstructure of $\sqrt{5}\times\sqrt{5}$ (in unit of nearest neighbor iron-iron distance), separating square blocks of four iron-atoms. The distinct geometry and the magnetic frustration thereafter lead to a unique antiferromangetic (AFM) ground state pattern, i.e. the block-spin AFM\cite{245_neutron,245_dft}.

  The presence of the ordered vacancy patterns alters the symmetry of the crystal lattice, and therefore its effect is far beyond the simple rigid-band shifting model. The new crystal symmetry modulates the electronic structure of the vacancy-free $A$Fe$_2$Se$_2$ (or 122) phase, and the iron vacancies introduce extra scattering centers that further modify the electronic states. Such effect of electronic structure reconstruction is absent in the vacancy disordered compounds\cite{PhysRevLett.109.147003,PhysRevLett.107.257001}; thus, the electronic structure of these vacancy-ordered phases can be completely different one from another, although their chemical composition may look fairly close. It therefore leads to a question that among all the possible vacancy-ordered phases, which one is the actual parent phase for the superconductivity. As the 245 phase seems to be present universally in these superconducting samples, it was proposed that the pairing was mediated by AFM fluctuation in the doped 245 compound\cite{245_neutron,PhysRevLett.107.167001}. This argument was challenged by a phase-separation scenario that the 245 phase and the superconducting phase are spatially separated. The phase separation argument has been supported by angle resolved photo-emission spectroscope (ARPES)\cite{PhysRevLett.106.187001,Zhang:2011qy} and scanning tunneling microscope (STM) or transmission electron microscopy (TEM) experiments\cite{STM_phase_separation,PhysRevLett.109.057003,PhysRevB.83.140505}, which suggest the vacancy concentration is much smaller and are disordered in the superconducting phase\cite{epl_zhou_zhang}. 

  Very recently, there has been an experimental report on $T_c$=33K superconducting iron selenides, which consists of superconducting bricks embedded in a background of 245 phase. The bricks have chemical component K$_{0.5+x}$Fe$_{1.75+y}$Se$_2$.  In these bricks which take up about 20\% of the volume, it is found that $x\approx$0.15 and $y\approx$0.01. This raises an interesting possibility that the vacancy-ordered K$_{0.5}$Fe$_{1.75}$Se$_2$ (278 phase hereafter), may be a parent compound of the superconducting phase\cite{278_expt}. The authors presented the preliminary verification on this possible parent phase by conducting self-consistent measurements of chemical compositions and the atomic imaging through scanning tunneling. Therefore, it will be interesting and urgent to provide first principles calculations and theoretical studies on this 278 phase.  Note that unlike the 245 phase, the single crystal 278 phase has not been discovered yet, and the phase form only "spider-web" like filaments.  

  In this article, we present our latest first principles results of the 278 phase. We examine the structural distortions induced by the iron vacancies; calculate its band structure and density of states, and compare them with the 122 phase. We present and compare its bare electronic susceptibility and nesting functions at zero and 0.1e doping levels. From these calculations, we conclude that the ground state for the 278 phase is stripe-like AFM and metallic. Upon 0.1e electron doping, which quickly suppresses the long range AFM order, the two-dimensionality of the Fermi surface sheets are greatly enhanced, resulting in a better nesting behavior.

{\it Method}-
  In the current study, we employed density functional theory (DFT) using plane-wave basis as implemented in Vienna Abinit Simulation Package (VASP) \cite{vasp, vasp_paw}. The ion-valence electron interactions were modeled using projected augmented wave (PAW) method; and the Perdew, Burke, and Ernzerhof flavor of exchange-correlation functional were chosen\cite{PBE}. A 480 eV energy cut-off for the plane-wave basis and a 6$\times$6$\times$3 $\Gamma$-centered K-grid ensures the total energy converges less than 1 meV/atom. The lattice constants and internal atomic positions were fully optimized until forces on individual atoms are smaller than 1 meV/\AA\ and total stress less than 0.05 kB. The density of states (DOS) calculations were performed on a much denser K-grid of 24$\times$24$\times$12 and tetrahedra method.

  The obtained band structure were then fitted to a tight-binding hamiltonian of 70 Fe-3d orbitals using maximally projected wannier function method\footnote{To do this, we used the wannier90 code without performing the optimization procedure. As pointed out in Ref. \cite{unfold_wannier}, the optimization procedure may alter the symmetry of the wannier orbitals, leading to potential problems in the band structure unfolding process later on.}. The resulting hamiltonian was then used to interpolate the electronic states on a dense $\Gamma$-centered $100\times 100\times 100$ K-grid. The bare electronic susceptibility $\chi_0$ is then calculated using 
\begin{align*}
\chi_0(\mathbf{q})=&\frac{1}{N_{\mathbf{k}}}\sum_{mn}\sum_{\mu\nu\mathbf{k}}\frac{\langle m\vert\mu\mathbf{k}\rangle\langle\mu\mathbf{k}\vert n\rangle\langle n\vert\nu\mathbf{k+q}\rangle\langle\nu\mathbf{k+q}\vert m\rangle}{\epsilon_{\nu\mathbf{k+q}}-\epsilon_{\mu\mathbf{k}}+i0^{+}} \\
 &\times\left[f(\epsilon_{\mu\mathbf{k}})-f(\epsilon_{\nu\mathbf{k+q}})\right]
\end{align*}

where $\epsilon_{\mu\mathbf{k}}$ and $f(\epsilon_{\mu\mathbf{k}})$ are the band energy (measured from $E_F$) and occupation number of $\vert \mu\mathbf{k}\rangle$, respectively; $\vert n\rangle$ denotes the $n^{th}$ wannier orbital; and $N_{\mathbf{k}}$ is the number of the $\mathbf{k}$ points used for the irreducible Brillouin zone (IBZ) integration. It is worthy noting that one should not make the usual assumption that $\langle \mu\mathbf{k}\vert n\rangle=1$, as pointed out by S. Graser {\it et al.} \cite{1367-2630-11-2-025016}. The information is also used to calculate the nesting function using
  $$f(\mathbf{q})=\frac{1}{N_{\mathbf{k}}}\sum_{\mathbf{k}}\delta(\epsilon_{\mu\mathbf{k}}) \delta(\epsilon_{\nu\mathbf{k+q}})$$.

\begin{figure}[ht]
 \centering
 \subfigure[Unit Cell]{
   \scalebox{0.15}{\includegraphics{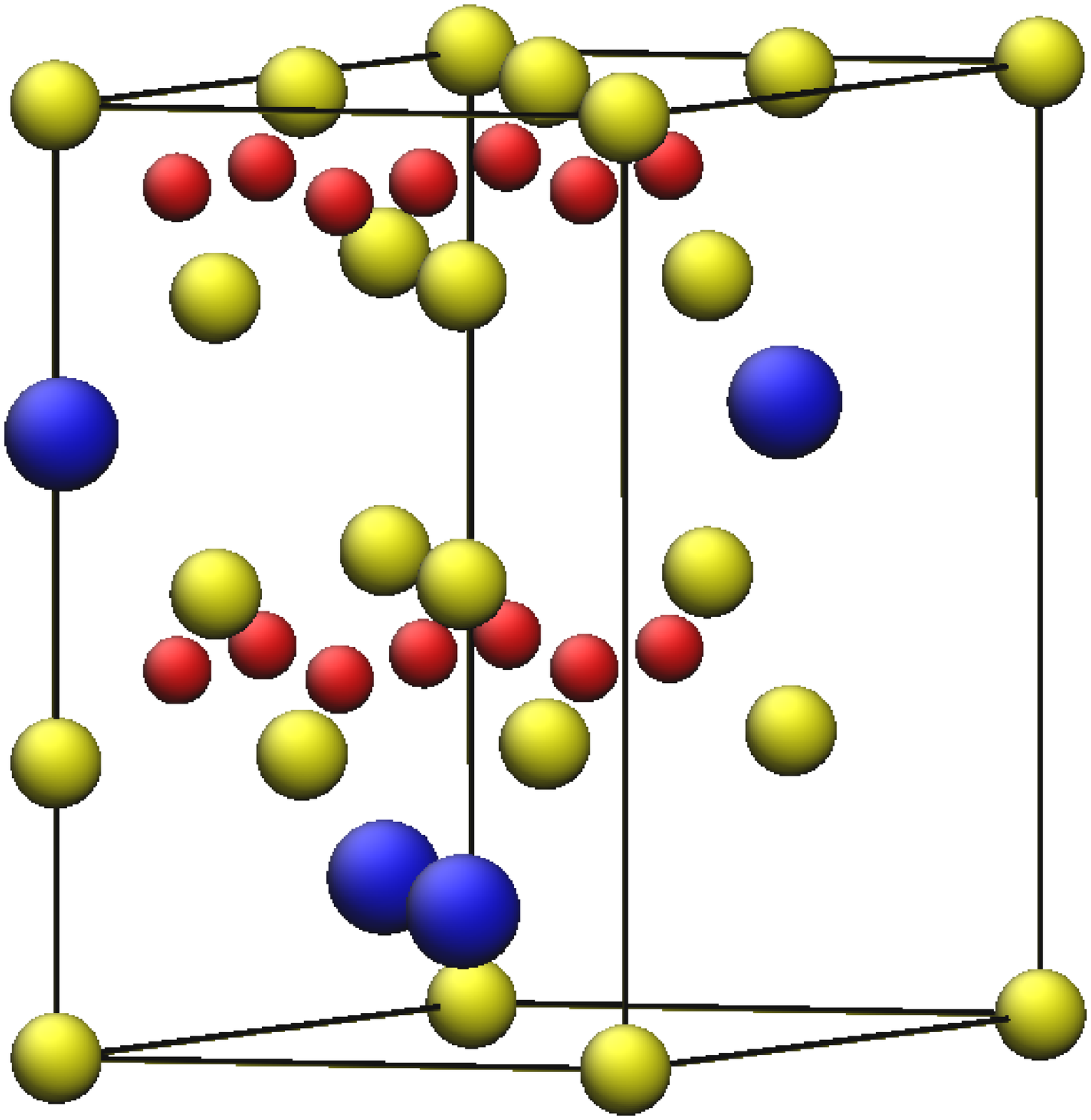}}
   \label{fig:geo_3d}
 }
 \subfigure[Fe-Plane]{
   \scalebox{0.15}{\includegraphics{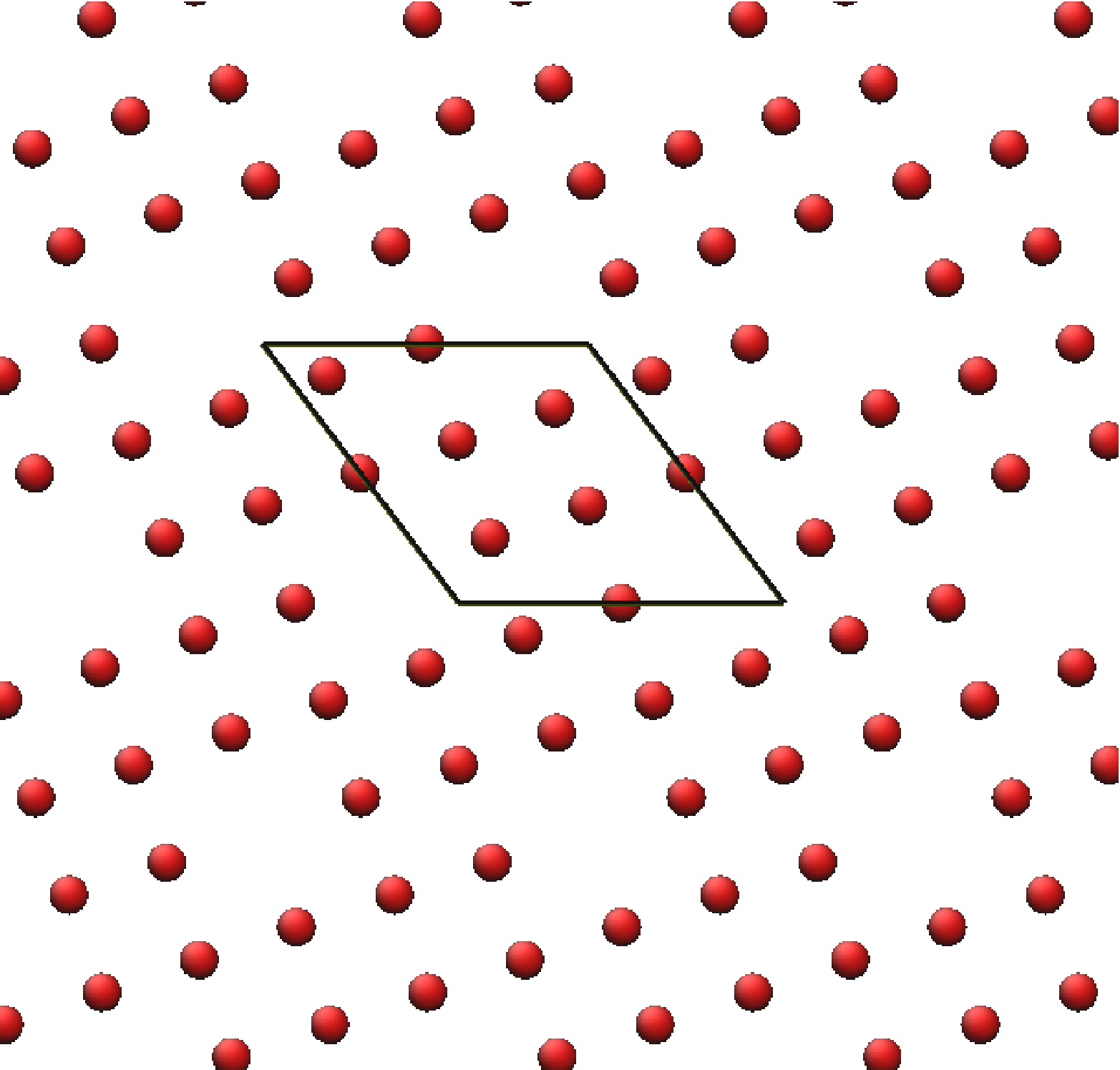}}
   \label{fig:geo_fe}
 }
 \subfigure[BZ]{
   \rotatebox{270}{\scalebox{0.2}{\includegraphics{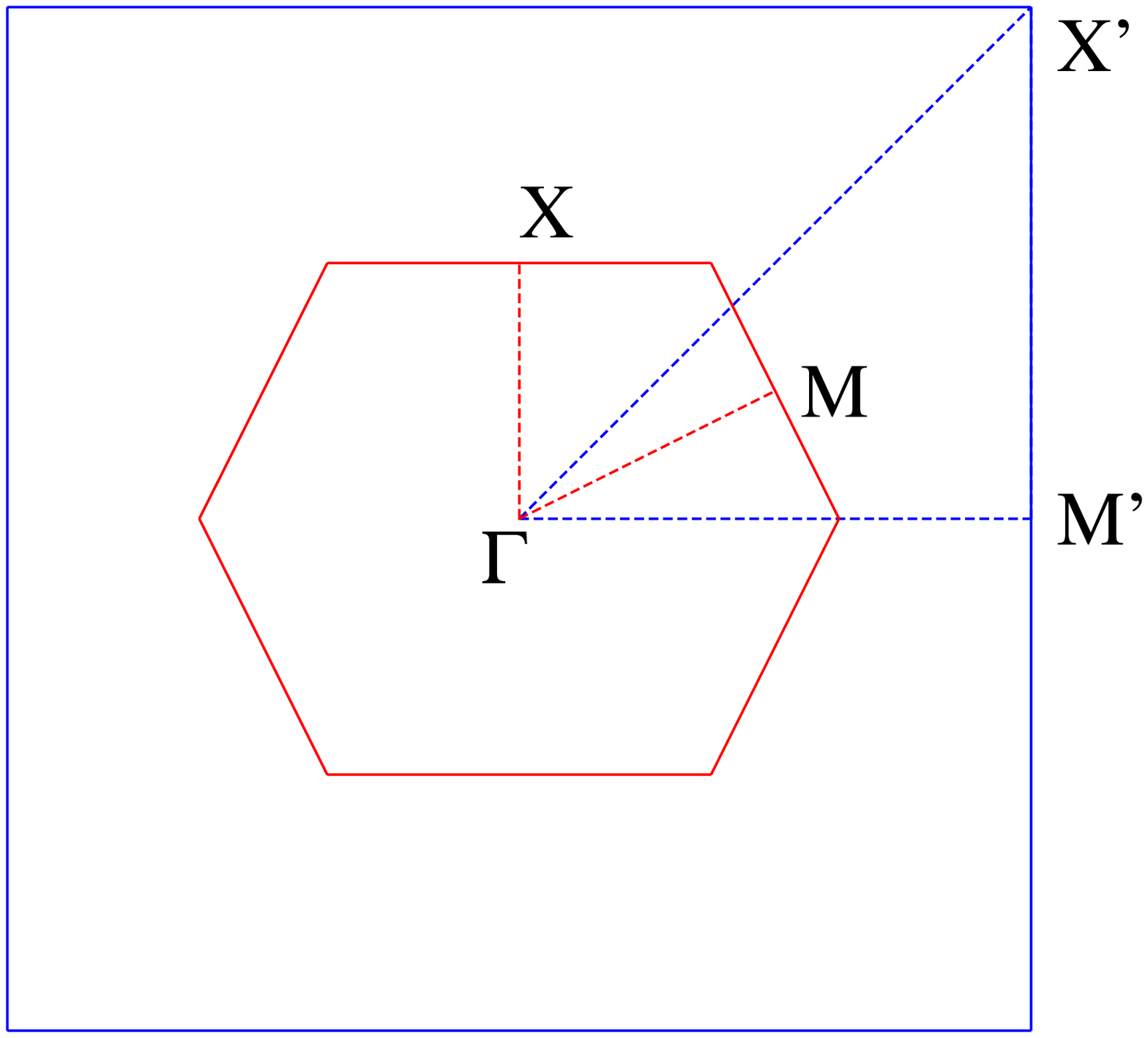}}}
   \label{fig:geo_bz}
 }
 \caption{(a) The unit cell of 278 phase. The red atoms are iron; the blue atoms are potassium; and the yellow atoms are selenium. (b) The iron-plane showing the ordered vacancies and the choice of lattice vectors as in (a). It is obvious that the two lattice vectors are both $\sqrt{10}$ (in unit of nearest neighbor Fe-Fe distance). (c) The first BZ (solid lines) and IBZ (dotted lines) of the 122 phase (the blue lines) and the 278 phase (the red lines). Both $M'$ and $X'$ in the 122 IBZ are $X$ in the 278 IBZ. \label{fig:geometry}}
\end{figure}

{\it Results and discussion}-
  Figure \ref{fig:geo_3d} and \ref{fig:geo_fe} show the unit cell we chose to perform our calculations. While the iron vacancy order is crucial to the electronic structure of these compounds, the effect of potassium order seems to be negligible except for the resulting chemical potential shifting, as shown in previous studies\cite{234_dft,245_dft}. Therefore, without losing generality, we chose a unit cell with high symmetry where occupied potassium sites are closest to the iron-vacancy sites. Figure \ref{fig:geo_bz} compares the IBZ of a 122 phase unit cell (enclosed by blue dotted lines) and the one of a 278 phase unit cell (enclosed by red dotted lines). It is worthy noting that the area of the first Brillouin zone (BZ) for the 122 phase (the blue square) is four times the one for the 278 phase (the red hexagon).
  
  The optimized lattice constants for the 278 phase are $a=8.3813$ \AA\ and $c=13.9846$ \AA\ (Fig. \ref{fig:geo_fe}), respectively. Previous density functional studies of iron-pnictides and iron-selenides severely suffer from a $c$-collapse problem that optimized lattice constant $c$ are significantly smaller (beyond usual DFT limit of 5\%) than the experimental value. This particular problem is in fact absent in the 278 phase if one assumes that the 278 phase $c$ value should not be too far from the 122 phase (14.0367 \AA). The $\gamma$ angle formed by $a$ and $b$ lattice vectors is 127.12$^{\circ}$, which is slightly larger than 126.87$^{\circ}$. The latter is the ideal value of $\gamma$ if the iron square lattices are perfectly preserved. Therefore, the square lattice of iron is slightly distorted by less than 0.5\%. Furthermore, the optimized lattice constants can be roughly translated into $a'=3.748$\AA\ in the 122 phase, which is 4.3\% smaller than the experimental value, suggesting a contraction of lattice constants due to the vacancy ordering. The presence of the ordered vacancies also introduces further distortions of the atomic positions within one unit cell. The seven iron-sites in one unit cell splits into three nonequivalent classes: the four iron atoms that are nearest neighbors of the iron-vacancy (Fe$^{\mathrm{I}}$); the iron site that is in between two neighboring iron-vacancies (Fe$^{\mathrm{II}}$); and the rest two iron atoms (Fe$^{\mathrm{III}}$), and their fractional heights are 0.2488, 0.2463 and 0.2523, respectively. In addition, the Fe$^{\mathrm{I}}$ atoms slightly move towards the nearest vacancy site while both Fe$_{\mathrm{II}}$ and Fe$_{\mathrm{III}}$ remains in square lattice position (so that they only displace in $z$-direction). The Fe-Se-Fe angles formed by the next-nearest-neighbor iron-sites and the above selenium atom are in the range of 108.16$^{\circ}$ to 109.66$^{\circ}$, slightly larger than the one in the KFe$_2$Se$_2$ 122 phase (106.61$^{\circ}$).

\begin{figure}[ht]
 \centering
 \subfigure[Band Structure] {
   \rotatebox{270}{\scalebox{0.6}{\includegraphics{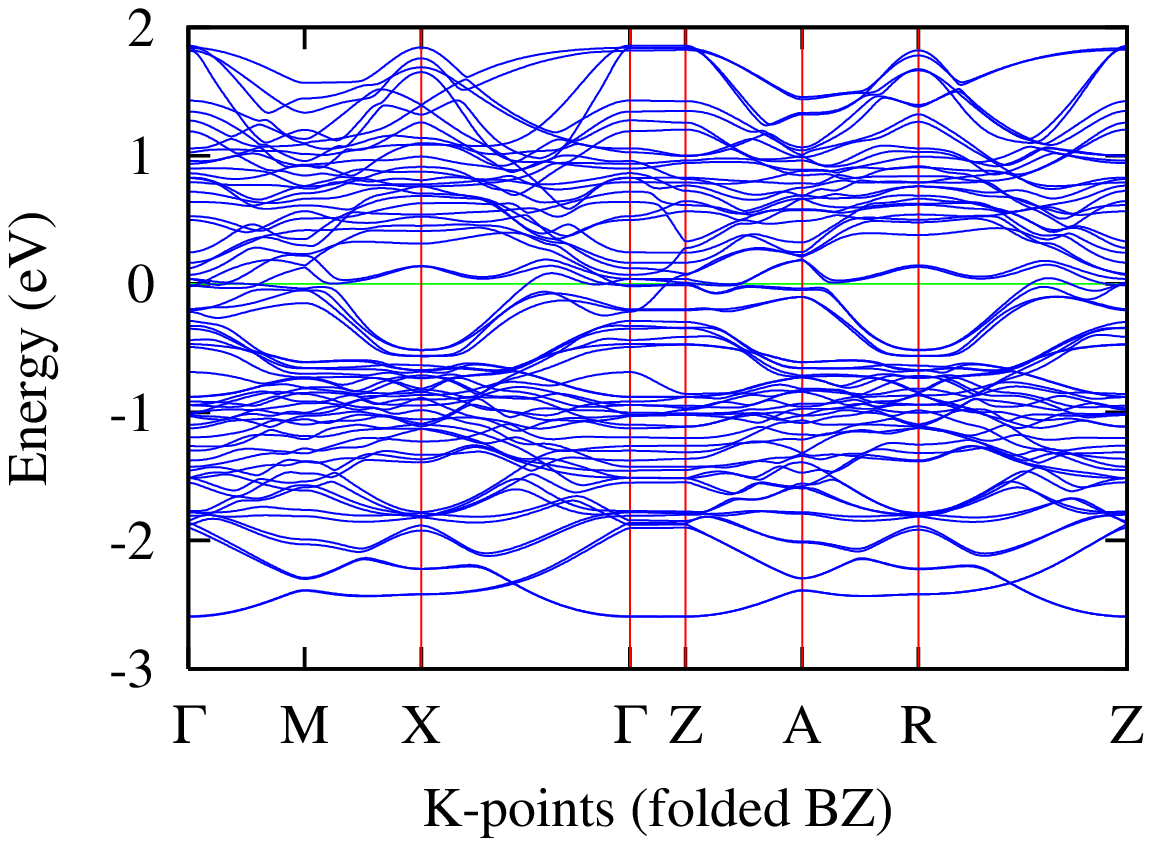}}}
   \label{fig:bs}
 }
 \subfigure[Density of States] {
   \rotatebox{270}{\scalebox{0.6}{\includegraphics{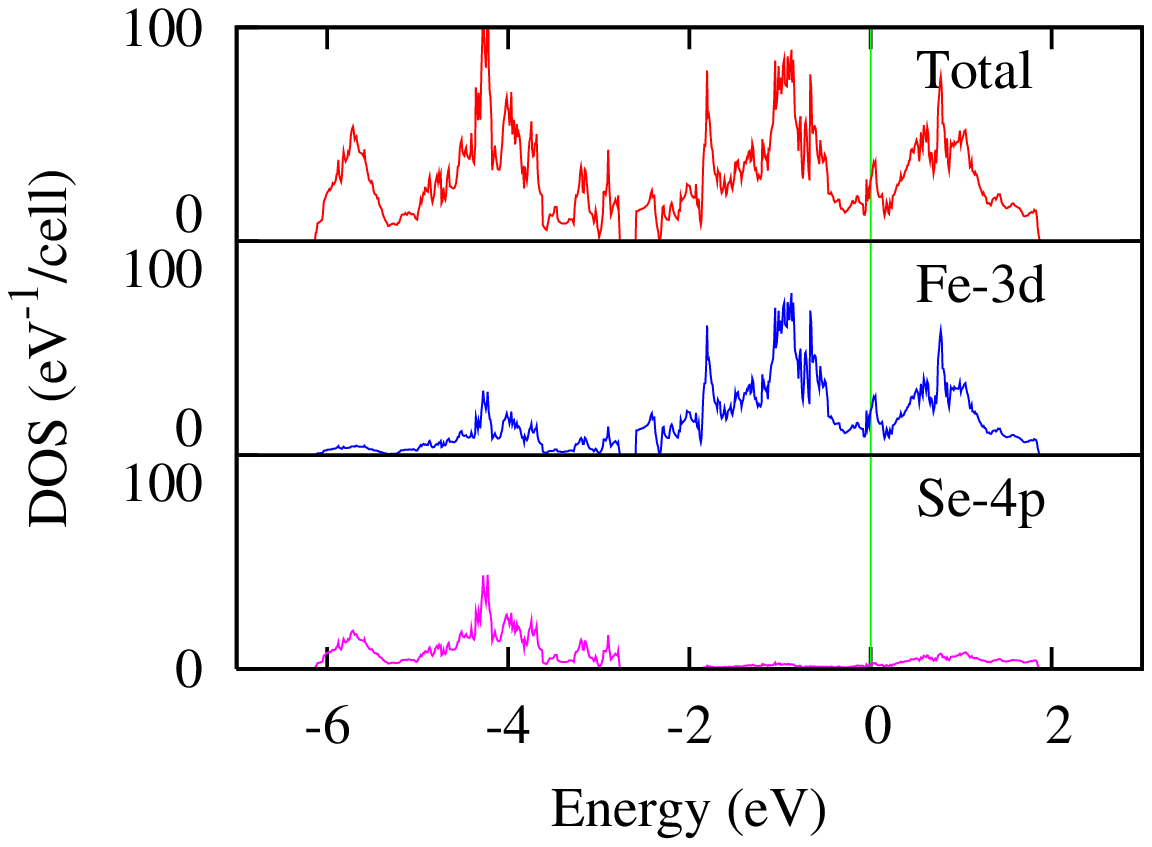}}}
   \label{fig:dos}
 }
 \caption{(a) Band structure and (b) density of states of 278 phase. Unlike other iron-pnictides or chalcogenides, the electronic states of 278 phase from $E_F$-3.0 eV to $E_F$+2.0 eV are almost exclusively from Fe-3d orbitals.\label{fig:bsdos}}
\end{figure}

\begin{figure}[ht]
 \subfigure[278 Unfolded] {
   \rotatebox{270}{\scalebox{0.6}{\includegraphics{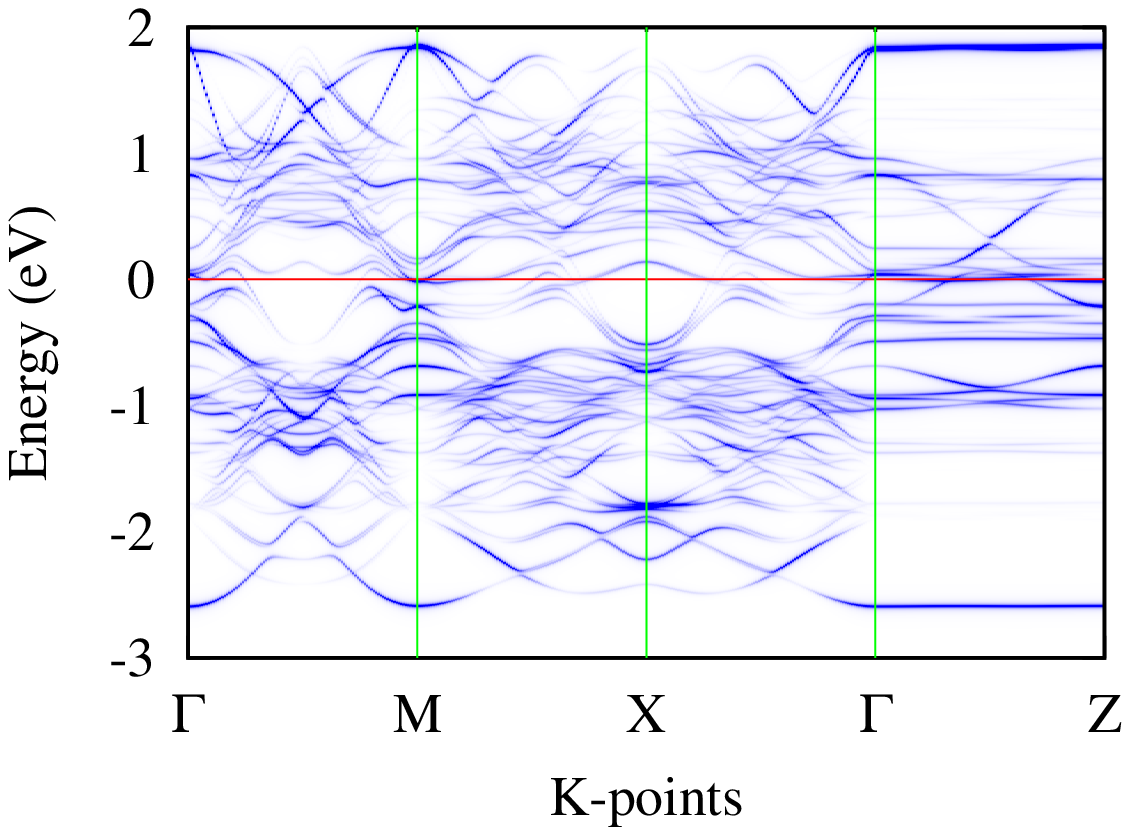}}}
   \label{fig:bs_278}
 }
 \subfigure[122] {
   \rotatebox{270}{\scalebox{0.6}{\includegraphics{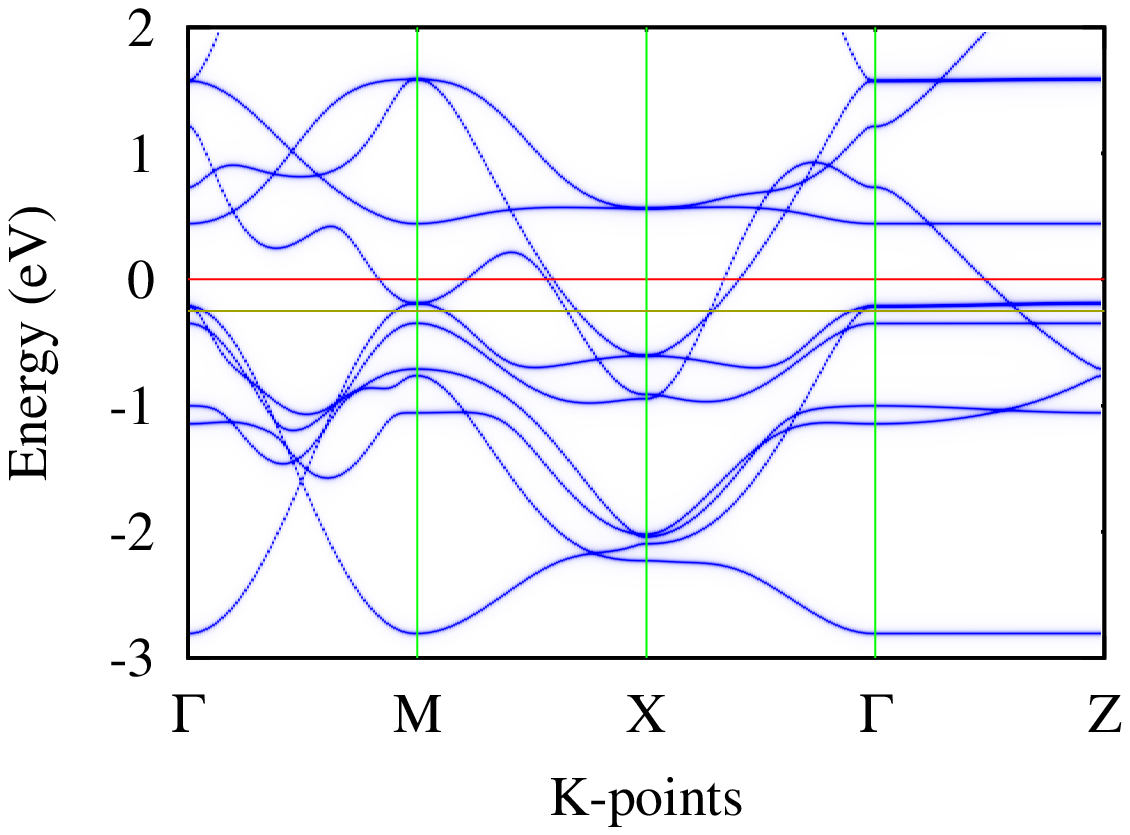}}}
   \label{fig:bs_122}
 }
 \caption{Band structure of (a) 278 phase in the unfolded BZ (122 phase BZ), and (b) 122 phase in 122 phase BZ. In both figures, the Fermi level $E_F$ is aligned at zero energy (red line), while in the 122 phase plot, an additional -0.25 eV line (corresponding to the doping level in the 278 phase) is shown.}
\end{figure}

  With the optimized crystal structure, we then study its electronic structure. Figure \ref{fig:bs} and \ref{fig:dos} show the band structure and density of states (DOS) of 278 phase. As the unit cell of the 278 phase is 8 times the 122 phase primitive cell (which is not orthogonal), its irreducible Brillouin zone (IBZ) as well as the band structure is heavily folded (for the IBZ folding, please refer to Fig. \ref{fig:geo_bz}). However, it can still be identified that three bands crosses the Fermi level, suggesting three Fermi surface sheets in the BZ. Unlike either 122 or 245 phase, the Fe-3d bands in the 278 phase are well separated from all other contributions, and dominates the electronic states from $E_F$-3.0 eV to $E_F$+2.0 eV as suggested in the projected DOS calculations (Fig. \ref{fig:dos}). Several flat bands can be identified close to $E_F$, from $\Gamma$ to M and from Z to A, which contributes the DOS peak at $E_F$. It is worthy noting that the flat band contribution to conductivity is almost negligible due to the large effective mass. Thus, the pseudogap-like feature from $E_F$-1.0 eV to $E_F$+1.0 eV may be consistent with the "bad-metal" behavior observed in the experiments.

  As the electronic states near $E_F$ are almost exclusively contributed by the Fe-3d orbitals, we can easily fit its band structure within $E_F$-3.0 eV to $E_F$+2.0 eV to a tight-binding hamiltonian using maximally projected wannier orbitals. The original 278 phase band structure is then unfolded to the 122 phase IBZ using the method introduced by W. Ku {\it et al.}\cite{unfold_wannier}. The resulting unfolded band structure is shown in Fig. \ref{fig:bs_278} in comparison with the 122 phase band structure (Fig. \ref{fig:bs_122}). Due to the strong structural distortion and iron vacancies, the band structure of the 278 phase is significantly smeared out, although some feature of the latter can still be identified from the unfolded 278 band structure. In particular, both the upper and lower band limit of Fe-3d orbitals seem to be shifted $\sim 0.25$ eV, meaning the Fermi level $E_F$ of the 278 phase is shifted $\sim -0.25$ eV. It is worthy noting that this shifting of $E_F$ corresponds to $\sim 0.5\vert e\vert$ hole doping in the 122 unit cell, or K$_{0.75}$Fe$_2$Se$_2$ in chemical formula. Similar effective doping feature was also previously identified in the 245 phase\cite{PhysRevLett.107.257001}. One can also identify some other characteristic rigid-band shifting in the unfolded 278 phase band structure, especially along $\Gamma$-$Z$ since the direction is mostly unaffected by the iron-vacancies whose effect is primarily in-plane.

  Features beyond rigid-band shifting can also be identified in the unfolded band structure. First of all, the additional scattering centers caused significant band smearing and splitting in the regions around $E_F-1.0$ eV and $E_F+1.0$ eV, similar to the unfolded band structure of the 245 phase\cite{PhysRevLett.107.257001}. Secondly, The -0.25 eV Fermi level shifting allows two hole-like Fermi surface sheets emerge around $\Gamma$ in the 122 phase, as one can clearly identify two hole like band dispersion cross the -0.25 eV line in Fig. \ref{fig:bs_122}. This feature, however, is absent in the 278 phase band structure. One another shallow hole-like band crossing, however, can be identified around $X$ in the 278 phase unfolded band structure, which is clearly missing in the rigid-band shifted 122-phase. Finally, the bands around $E_F$ are extremely smeared out except from $\Gamma$ to $Z$, suggesting that the electron transport suffers from strong scattering in the $a$-$b$ plane and may exhibit a low conductivity. Due to the complexity of the band structure, it is rather difficult to conclude the type of the major carrier, but both electron and hole seems to play roles.

\begin{figure}[ht]
 \centering
 \subfigure[Undoped susceptibility] {
   \rotatebox{270}{\scalebox{0.2}{\includegraphics{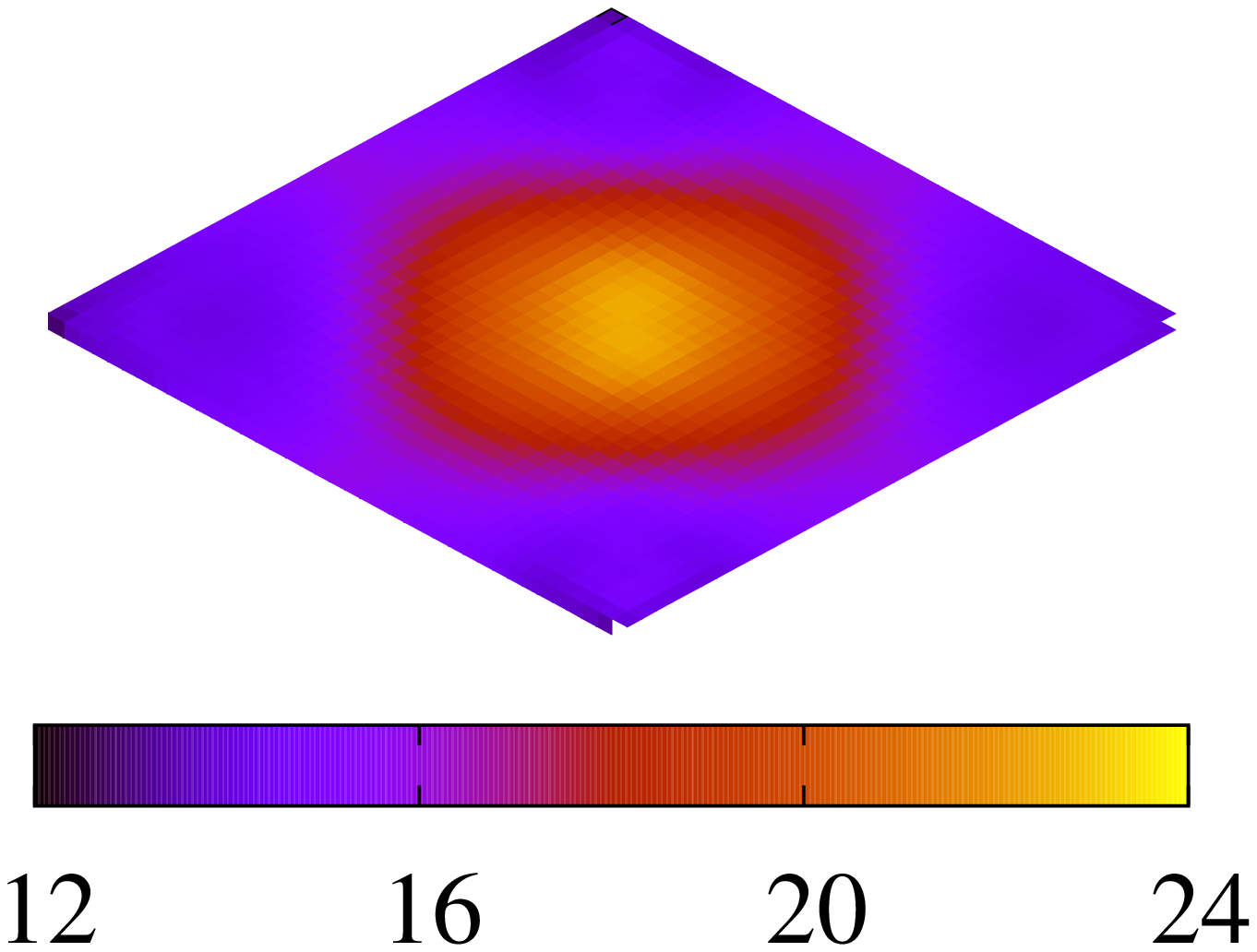}}}
   \label{fig:chi_parent}
 }
 \subfigure[Doped susceptibility] {
   \rotatebox{270}{\scalebox{0.2}{\includegraphics{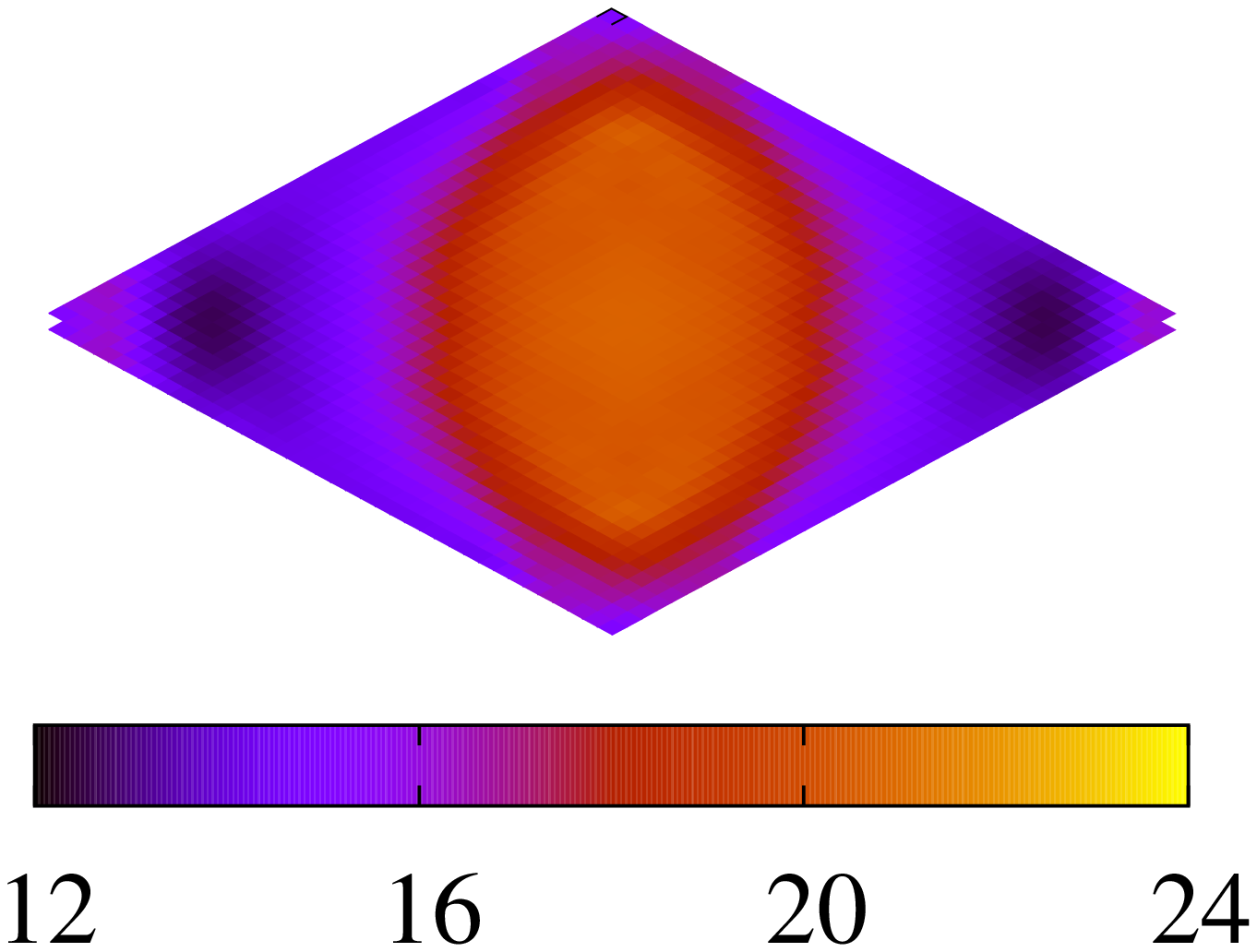}}}
   \label{fig:chi_doped}
 }
 \subfigure[Undoped nesting] {
   \rotatebox{270}{\scalebox{0.2}{\includegraphics{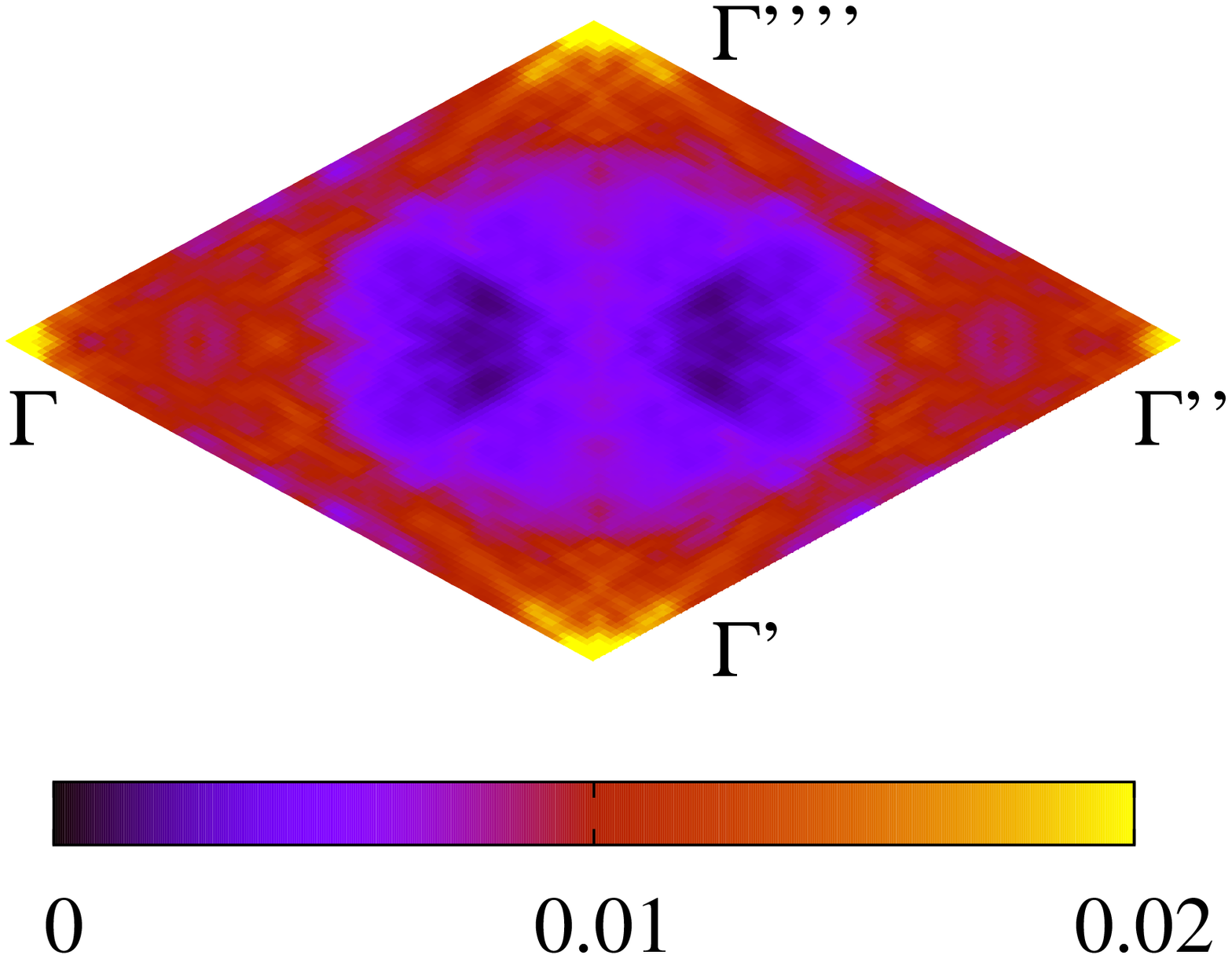}}}
   \label{fig:nest_parent}
 }
 \subfigure[Doped nesting] {
   \rotatebox{270}{\scalebox{0.2}{\includegraphics{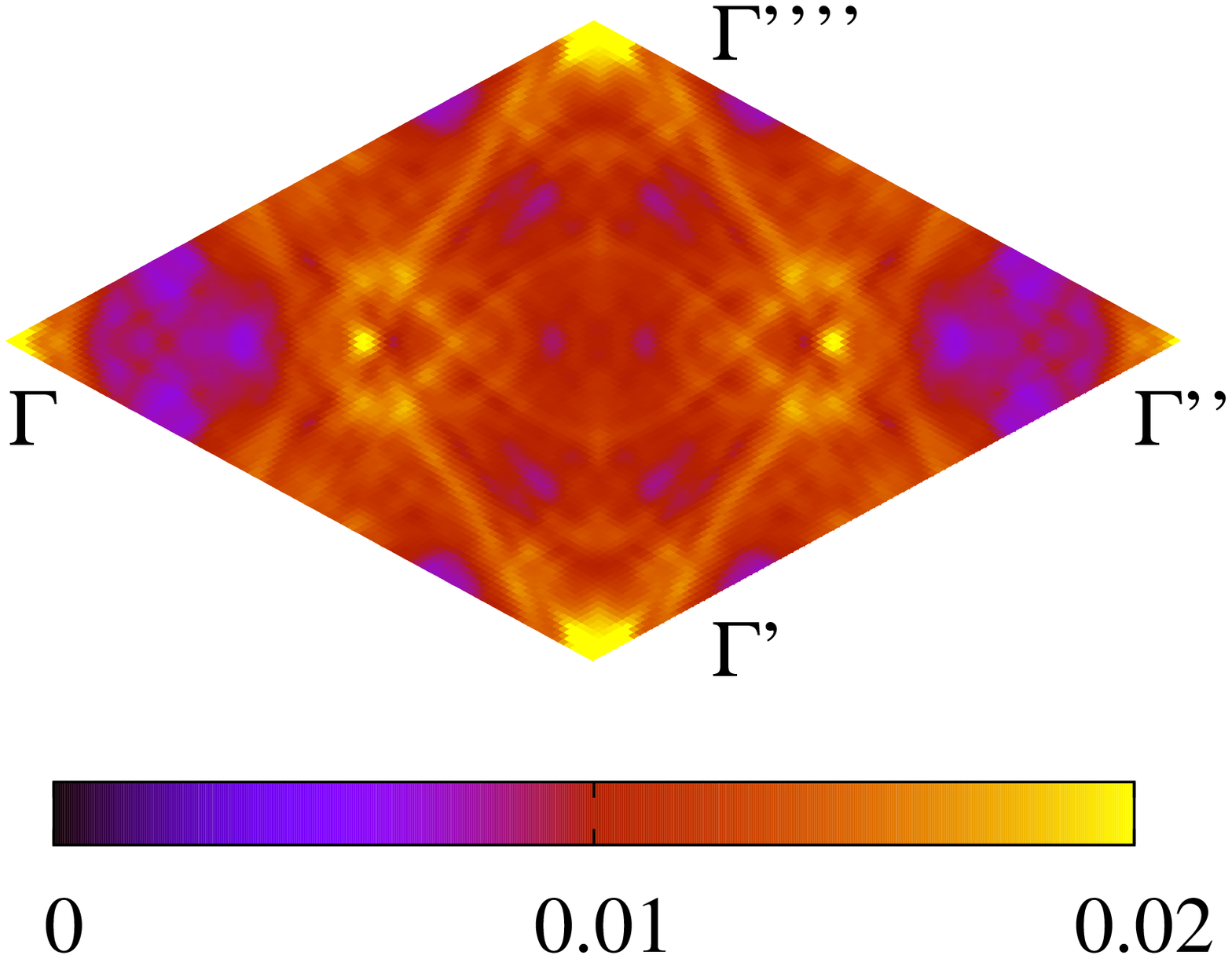}}}
   \label{fig:nest_doped}
 }
 \caption{(a-b) Bare electron susceptibility $\chi_0$ and (c-d) Nesting function $f$ of parent 278 phase (a, c) and 0.1e electron doped 278 phase (b,d). The plots are made in a full BZ with $\Gamma$ at four corners. }
\end{figure}

  The bare electronic susceptibility $\chi_0$ is also calculated using the maximally projected wannier function hamiltonian. Fig. \ref{fig:chi_parent} and \ref{fig:chi_doped} show $\chi_0$ at $k_z=0$ plane for both the parent 278 phase and 0.1e-doped phase, respectively. Previous calculations for BaFe$_2$As$_2$ and LaOFeAs\cite{PhysRevLett.101.057003,PhysRevB.81.214503} show significant enhancement at $(\pi, 0)$ in the two-iron (unfolded) BZ. For the parent 278 phase, a similar enhancement of $\sim$100\% can also be identified around $(\pi, \pi)$ in the folded BZ, suggesting an AFM long range order. In fact, mean-field calculations based on the extended $J_1-J_2$ model have proposed four possible ground state candidates\footnote{Private communication with Feng Lu {\it et al.}.}. By employing spin-polarized density functional calculations, we confirm that the ground state parent 278 phase is stripe-like order AFM, which is $\sim$150 meV/Fe lower than the non-magnetic phase and $\sim$30 meV/Fe lower than the second lowest AFM order. With 0.1e electron doping, the enhancement is greatly reduced and broadened, thus the long range AFM order may be destroyed at such doping level. 

\begin{figure}[ht]
 \subfigure[Parental 278] {
   \scalebox{0.15}{\includegraphics{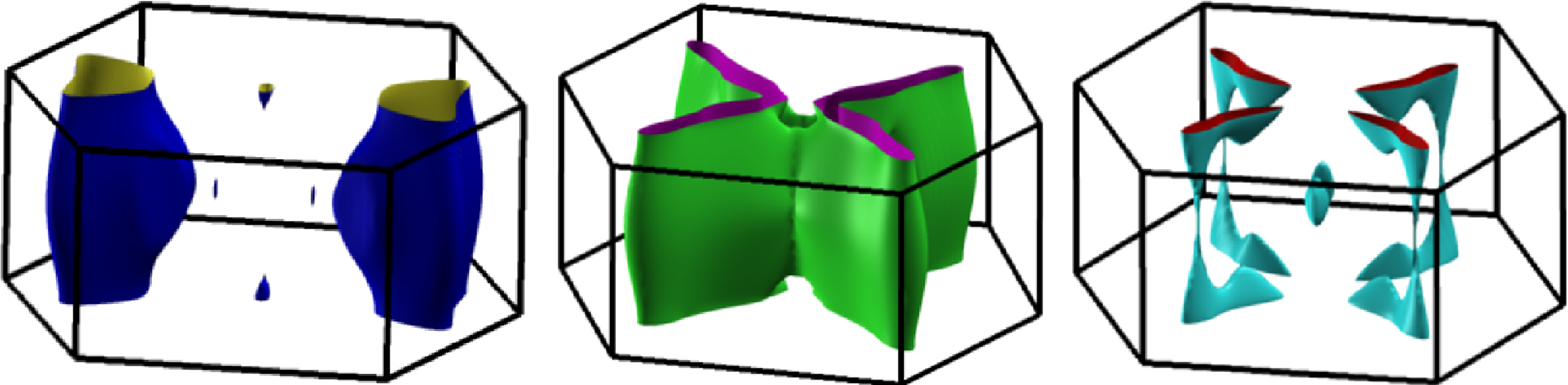}}
   \label{fig:fs_parent}
 }
 \subfigure[0.1 e doped] {
   \scalebox{0.15}{\includegraphics{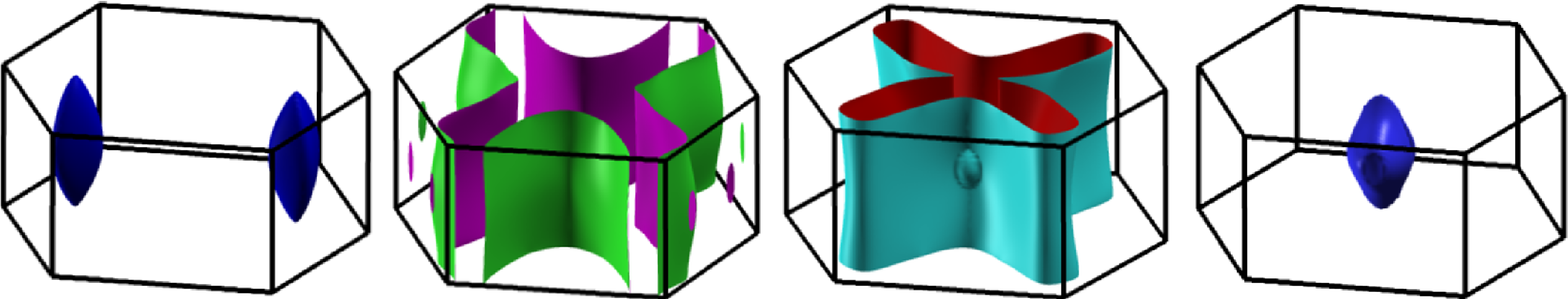}}
   \label{fig:fs_doped}
 }
 \caption{Fermi surface sheets of (a) parent 278 phase, and (b) 0.1e electron-doped 278 phase. The $\Gamma$ point is at the center of each plot.\label{fig:fs}}
\end{figure}

  Finally, we present the Fermi surfaces for both the parent compound and the 0.1e doped compound (Fig. \ref{fig:fs}). As pointed out previously, despite of the complex band structure of the 278 phase, only three bands across the Fermi level, forming three Fermi surface sheets. Upon electron doping, the sizes of the two hole pockets along $\Gamma$-M are dramatically reduced while the electron pockets are greatly enlarged. Apart from that, a fourth sheet emerges around $\Gamma$. Furthermore, the two-dimensionality of both electron pockets are significantly enhanced, leading to a better nesting behavior. This is also reflected in the nesting function plots (Fig. \ref{fig:nest_parent} and \ref{fig:nest_doped}), which show over twice enhancement of the nesting function around the center of BZ.

{\it Conclusion}
  In conclusion, we have performed first-principles study on vacancy-ordered K$_{0.5}$Fe$_{1.75}$Se$_2$ (278 phase). The ground state of the parent 278 phase is stripe-like AFM and metallic, whose magnetic order may be destroyed by electron/hole doping. The Fe-3d orbitals dominate the electron DOS near Fermi level, and both electron- and hole-like Fermi surfaces emerge in the BZ. The unfolded band structure shows limited features of a hole-doped 122 phase, but effects beyond rigid-band model must be taken into consideration to understand the electronic structure of the 278 phase. Finally, a 0.1e electron doping significantly enhances the two-dimensionality of its Fermi surfaces, resulting in a better nesting behavior. In our present calculations, electron correlations beyond local density approximation level have been neglected. A strong on-site Coulomb repulsion could alter the magnetic order and lead to Mott insulator, in light of the AFM Mott insulator in 245 compound.  The analysis is beyond the scope of the present paper and experiment synthesis of the 278 phase would shed new light on this issue.   

\begin{acknowledgments}
We would like to thank H.-H. Wen for sharing the experimental results and useful discussions. This work has been supported by the NSFC (No. 11274006) and the NSF of Zhejiang Province (No. LR12A04003). All calculations were performed at the High Performance Computing Center of Hangzhou Normal University College of Science.
\end{acknowledgments}

\bibliography{K278}
\end{document}